\documentclass[a4paper,11pt]{article}
\usepackage{pos}

\title{Equation of state and Taylor expansions at nonzero isospin chemical potential}
\ShortTitle{EoS and Taylor expansions at nonzero isospin chemical potential}

\author*[a]{Bastian B. Brandt}
\author[b]{Francesca Cuteri}
\author[a]{Gergely Endr\H{o}di}

\affiliation[a]{Institute for Theoretical Physics, University of Bielefeld, D-33615 Bielefeld, Germany}

\affiliation[b]{Institute for Theoretical Physics, Goethe University, D-60438 Frankfurt am  Main, Germany}

\emailAdd{brandt@physik.uni-bielefeld.de}
\emailAdd{endrodi@physik.uni-bielefeld.de}
\emailAdd{cuteri@itp.uni-frankfurt.de}

\abstract{We compute the equation of state of isospin asymmetric QCD at zero and non-zero temperatures using direct simulations of lattice QCD with three dynamical flavors at physical quark masses. In addition to the pressure and the trace anomaly and their behavior towards the continuum limit, we will particularly discuss the extraction of the speed of sound. Furthermore, we discuss first steps towards the extension of the EoS to small non-zero baryon chemical potentials via Taylor expansion.}

\FullConference{%
The 39th International Symposium on Lattice Field Theory,\\
8th-13th August, 2022,\\
Rheinische Friedrich-Wilhelms-Universität Bonn, Bonn, Germany
}

\newcommand{\ev}[1]{\left\langle #1 \right\rangle}

\begin{document}
\maketitle

\section{Introduction}

The equation of state constitutes the most relevant input for the phenomenological description
of physical systems in cosmology and astrophysics and is essential for the
hydrodynamical modeling of heavy-ion collisions. Most of these systems are dominated by
non-vanishing baryon density, but charge and strangeness densities can also contribute
significantly to the thermodynamics of the system. For some systems it is indeed the charge density
which plays the major role, such as in the case of an early Universe featuring sizeable lepton
flavour asymmetries~\cite{Wygas:2018otj,Middeldorf-Wygas:2020glx,Vovchenko:2020crk}.
In all of these cases, however, the EoS has to be known in the full three-dimensional parameter
space to allow for a full description of these pysical systems. The computation of the EoS
in most of the parameter space is hampered by the infamous sign problem at nonzero density.

In lattice QCD, we are typically working in the grand canonical ensemble, where the densities
are traded for the respective chemical potentials. Since in QCD the individual quark densities
are conserved, we are free to choose a suitable chemical potential basis. In the $N_f=2+1$ setup,
which is sufficient to capture the main dynamics for temperatures around the thermal transition
temperature, a convenient basis is given by
\begin{equation}
    \label{eq:chem-base}
    \mu_u=\mu_L+\mu_I \,, \qquad \mu_d=\mu_L-\mu_I \qquad \text{and} \quad \mu_s \,.
\end{equation}
This basis is ideally suited to distinguish between cases suffering from the complex action
problem and those which do not. As long as $\mu_L=\mu_s=0$, the case known as pure isospin
chemical potential, the action is real and the theory is amenable to Monte-Carlo
simulations~\cite{Son:2000xc,Kogut:2002tm,Kogut:2002zg}. A few years ago we started
the first dedicated program to extract the properties of QCD at non-zero isospin chemical
potential at the physical point with controlled systematics. This includes a detailed study
of the phase diagram~\cite{Brandt:2017oyy,Brandt:2018omg}, for which the presence
of a superconducting BCS phase at large $\mu_I$ is still an open
question~\cite{Brandt:2019hel,Cuteri:2021hiq}, as well as the extraction of the EoS.
First accounts of the results for the latter have already been presented for
zero~\cite{Brandt:2018bwq} and
nonzero~\cite{Vovchenko:2020crk,Brandt:2017zck,Brandt:2018wkp,Brandt:2021yhc}
temperatures. Here we will discuss the status of the extraction of the
EoS.

A related observable, which has become very prominent in the modelling of the EoS
for neutron stars in the past decade, e.g.~\cite{Tews:2018kmu,Annala:2019puf,Marczenko:2022jhl},
is the speed of sound $c_s$. In this proceedings article, we will discuss the extraction
of the speed of sound from the EoS at zero and non-zero temperature and show the first
results for the speed of sound obtained in the pion condensed phase from first principles in full
QCD at the physical point. In particular, we will see that for high isospin chemical potentials $c_s$ exceeds the conformal
bound~\cite{Cherman:2009tw} of $c_s=1/\sqrt3$. This is the first time that such a behavior has been
observed in full QCD within lattice simulations. In light of this, it appears instructive to reconsider what counts as a natural parameter
region for $c_s$.

Eventually, we are interested in the full three-dimensional parameter space. Our simulation points
at nonzero isospin chemical potential provide a novel starting point to extract the EoS in this
parameter space at small $\mu_L$ and $\mu_s$, but large $\mu_I$ using indirect methods, giving
access to previously inaccessible regions in parameter space. We will discuss the first steps
towards the extension of our EoS to $\mu_L\neq0$ using the Taylor expansion
method~\cite{Gottlieb:1988cq}.

\section{Equation of state at pure isospin chemical potential}

In our study, we use $N_f=2+1$ flavours of improved rooted staggered quarks with two levels of
stout smearing and tuned to physical quark masses, as well as the tree-level Symanzik improved
gluon action. To enable simulations in the phase where charged pions condense, the BEC phase,
the simulations entail a regulator, the pionic source, controlled by a parameter $\lambda$.
In particular, we simulate at $\lambda\neq0$ and results are extrapolated to $\lambda=0$
using the improvement program described in Ref.~\cite{Brandt:2017oyy}. The relevant observable
for the extraction of the EoS is the isospin density
\begin{equation}
    \langle n_I \rangle = \frac{T}{V}\frac{\partial \log\mathcal{Z}}{\partial \mu_I} \,.
\end{equation}
For the definition and the
improvement of the $\lambda$-extrapolations we refer to Ref.~\cite{Brandt:2018bwq}.
In this and the following section we will only discuss results which have already been
extrapolated to $\lambda=0$.

\subsection{The EoS at vanishing temperature}

The extraction of the EoS at (approximately) vanishing temperature has been discussed in
Ref.~\cite{Brandt:2018bwq}. The full EoS can be obtained from the isospin density, which, upon integration over $\mu_I$, gives the pressure and the interaction measure,
\begin{equation}
    p(0,\mu_I) = \int_0^{\mu_I} d\mu'_I \, n_I(0,\mu'_I) \qquad \text{and} \qquad
    I(0,\mu_I)=-4p+n_I(0,\mu_I) \mu_I \,.
\end{equation}

\begin{figure}[t]
 \centering
 \includegraphics[width=.45\textwidth]{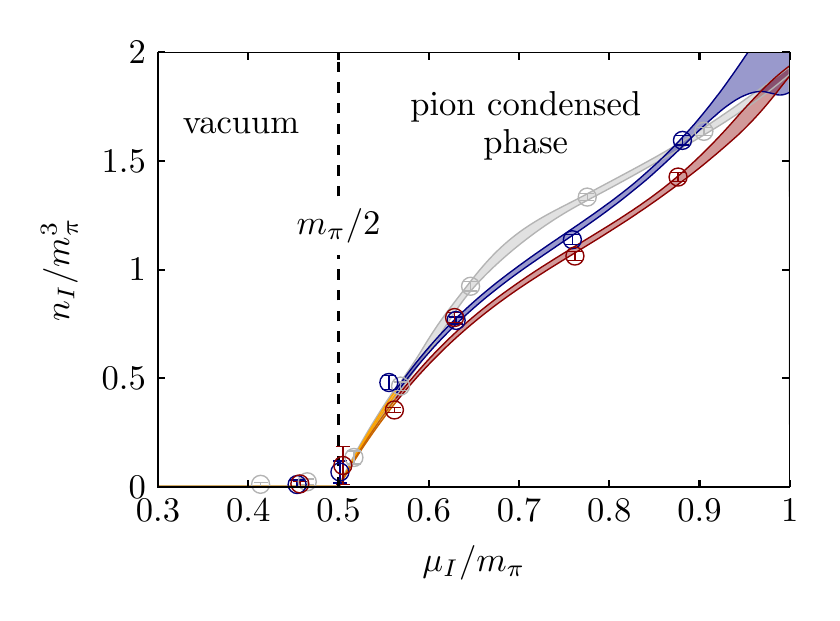}
 \includegraphics[width=.45\textwidth]{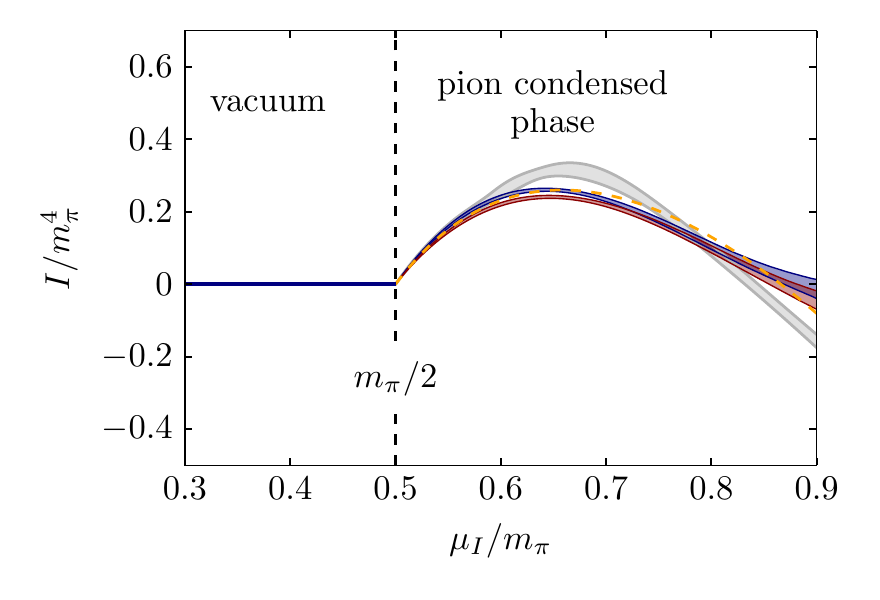}
 \caption{
 {\bf Left:} Simulation results for the isospin density $n_I$ vs. the isospin chemical potential
 $\mu_I$ at vanishing temperature together with the interpolation explained in the text. The red points
 and curve belong to the ensembles with $a\approx0.15$~fm and the blue to the ensembles with
 $a\approx0.22$~fm. The light gray results and curves belong to the old data at
 $a\approx0.29$~fm and a slightly smaller temperature published in Ref.~\cite{Brandt:2018bwq}.
 {\bf Right:} Results for the interaction measure obtained from the interpolation of the
 isospin density shown in the left panel. The mapping between colours and lattices is the same
 as in the left panel. The yellow dashed line is the result from chiral perturbation
 theory~\cite{Son:2000xc}.}
 \label{fig:eos-T0}
\end{figure}

The main problem when extracting the EoS are residual temperature artifacts, due to the
missing $T\to0$ extrapolations. In practice,
we simulate at $T\approx30$~MeV and correct for the residual
effects. Those are mostly prominent
in the vicinity of the transition to the BEC phase, see Fig.~\ref{fig:eos-T0}, where they lead
to non-vanishing values of the isospin density outside of the BEC phase. We correct these
artifacts by applying chiral perturbation theory at non-zero isospin chemical
potential~\cite{Son:2000xc} in the vicinity of the transition (see~\cite{Brandt:2018bwq}),
fitting the first two datapoints in the BEC phase and matching to cubic spline interpolations
of the remaining datapoints.
The spline interpolations are obtained as model-independently as possible by averaging over
spline fits with all possible nodepoint combinations, making use of Monte-Carlo methods~\cite{Brandt:2016zdy}.

Our study of the EoS at $T=0$ has been started on a comparably coarse lattice, with
$a\approx0.29$ fm~\cite{Brandt:2018bwq}. Here we
augment that study with two new sets of ensembles:
a set of $24^3\times32$ ensembles at a lattice spacing
of about $a\approx 0.22$~fm and a set of $32^3\times48$ ensembles at $a\approx 0.15$~fm.
The results for the interpolation of the isospin density are shown in the left panel of
Fig.~\ref{fig:eos-T0} and the resulting interaction measure in the right panel. The interaction
measure shows a clear sign for the presence of the BEC phase. It initially rises until it reaches
a maximum around $\mu_I/m_\pi\approx0.63$ to 0.65., from where it decreases until it
becomes negative around $\mu_I/m_\pi\approx0.9$, in good agreement~\cite{Adhikari:2019zaj} with chiral perturbation theory~\cite{Son:2000xc} (yellow dashed line).

\subsection{The EoS at nonzero temperature}

\begin{figure}[t]
 \vspace*{-2mm}
 \centering
 \begin{minipage}{.48\textwidth}
 \centering
 \includegraphics[width=\textwidth]{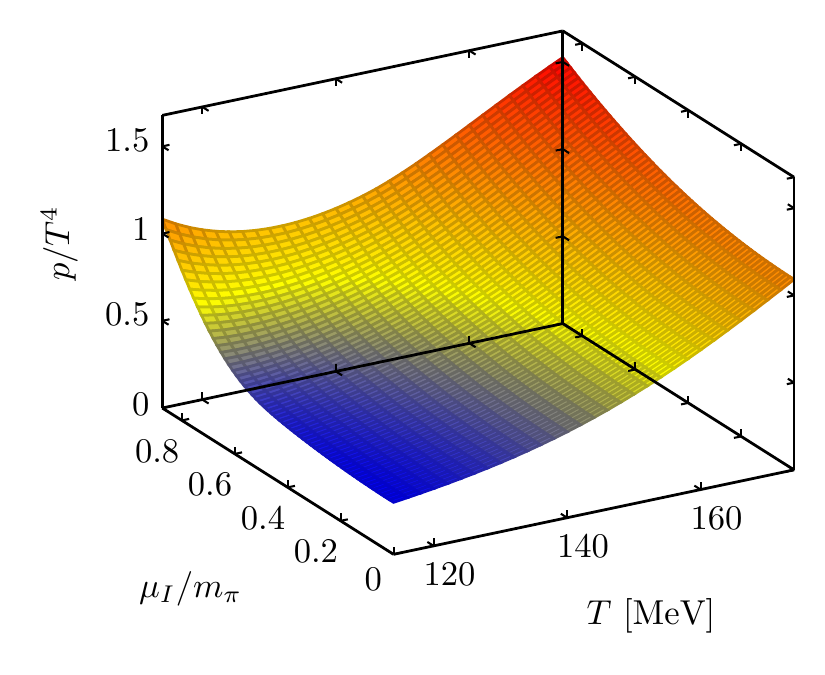} \\[-4.5mm]
 \includegraphics[width=\textwidth]{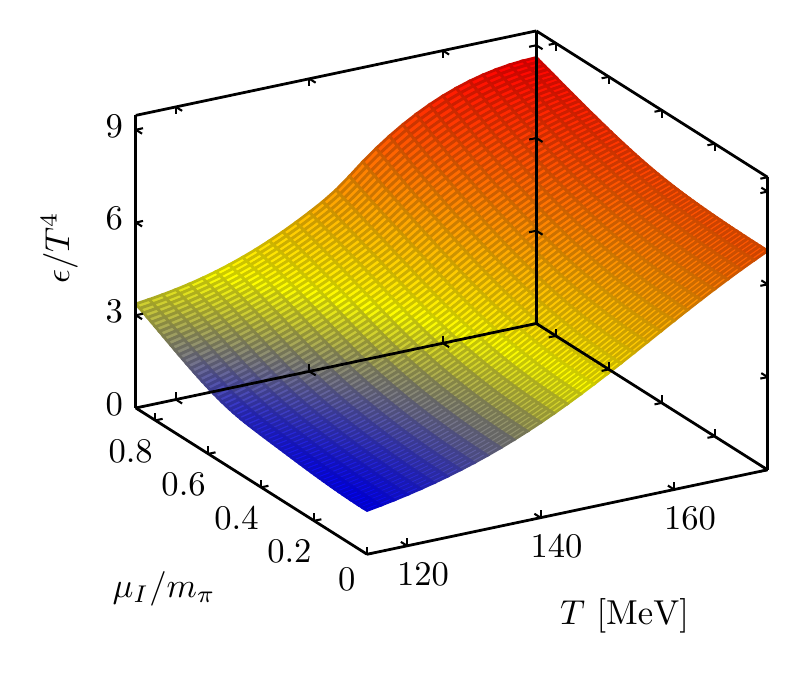}
 \end{minipage}
 \begin{minipage}{.48\textwidth}
 \centering
 \includegraphics[width=\textwidth]{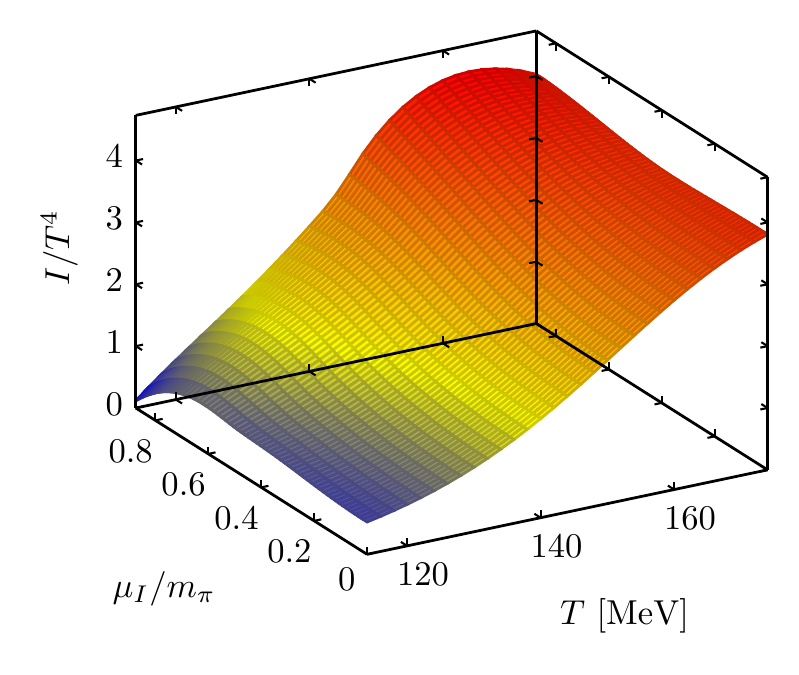} \\[-4.5mm]
 \includegraphics[width=\textwidth]{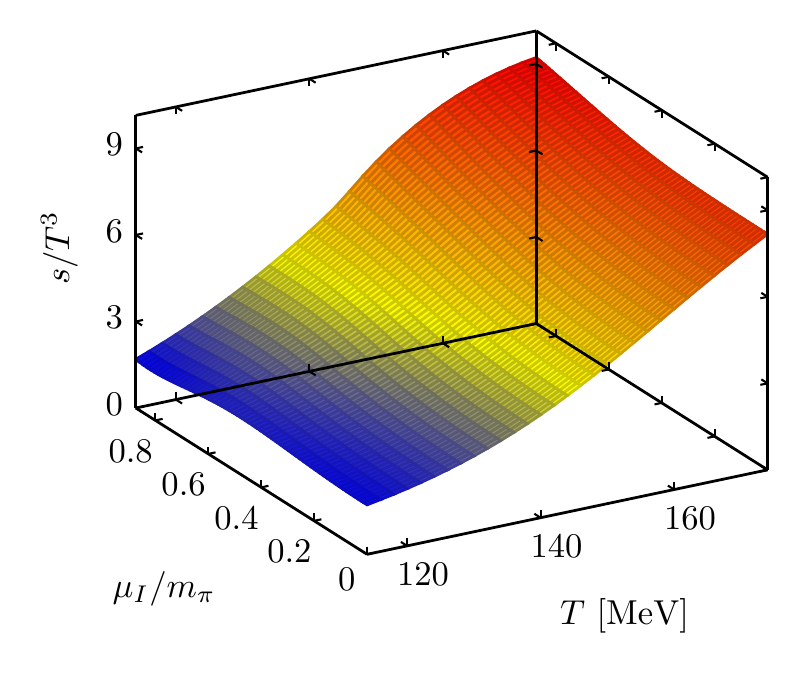}
 \end{minipage}
 \caption{
 Results for the pressure (top-left), the interaction measure (top-right), the energy density (bottom-left) and and the entropy density (bottom-right). The results have been obtained on our lattice with $N_t=8$ and for better visibility we do not show the uncertainties.}
 \label{fig:eos-nt8}
\end{figure}

At $T\neq0$, we can restrict ourselves to compute the modifications of the pressure
and the interaction measure due to the non-vanishing isospin chemical potential,
decomposing the two quantities as
\begin{equation}
 p(T,\mu_I) = p(T,0) + \Delta p(T,\mu_I) \quad \text{and} \quad I(T,\mu_I) = I(T,0) + \Delta I(T,\mu_I) \,.
\end{equation}
The $\mu_I=0$ contributions are available from the literature~\cite{Borsanyi:2013bia,HotQCD:2014kol}.
The modifications $\Delta p(T,\mu_I)$ and $\Delta I(T,\mu_I)$ can be computed, similar to the
$T=0$ case, from a, now two-dimensional, interpolation of the isospin density, again using a
model independent spline interpolation (see Refs.~\cite{Vovchenko:2020crk,Brandt:2021yhc}).
After computing the pressure and the interaction measure
from the interpolation, most of the other relevant thermodynamic quantities can be computed,
such as energy and entropy density, $\epsilon$ and $s$, for instance,
\begin{equation}
\label{eq:ene-ent}
 \epsilon = I+3p \quad \text{and} \quad s=\frac{\epsilon + p - \mu_I n_I}{T} \,.
\end{equation}

\begin{figure}[t]
 \vspace*{-2mm}
 \centering
 \begin{minipage}{.48\textwidth}
 \centering
 \includegraphics[width=\textwidth]{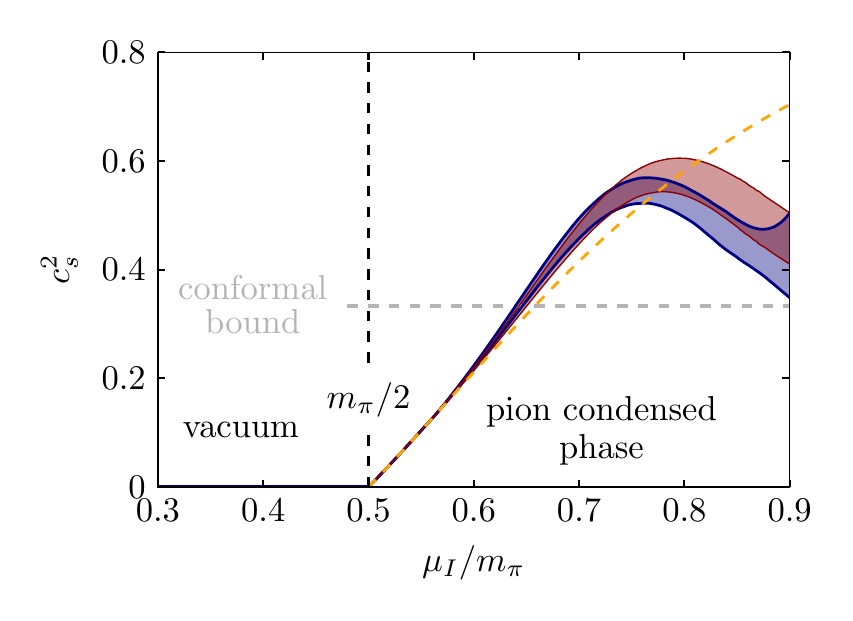} \\[4mm]
 \includegraphics[width=\textwidth]{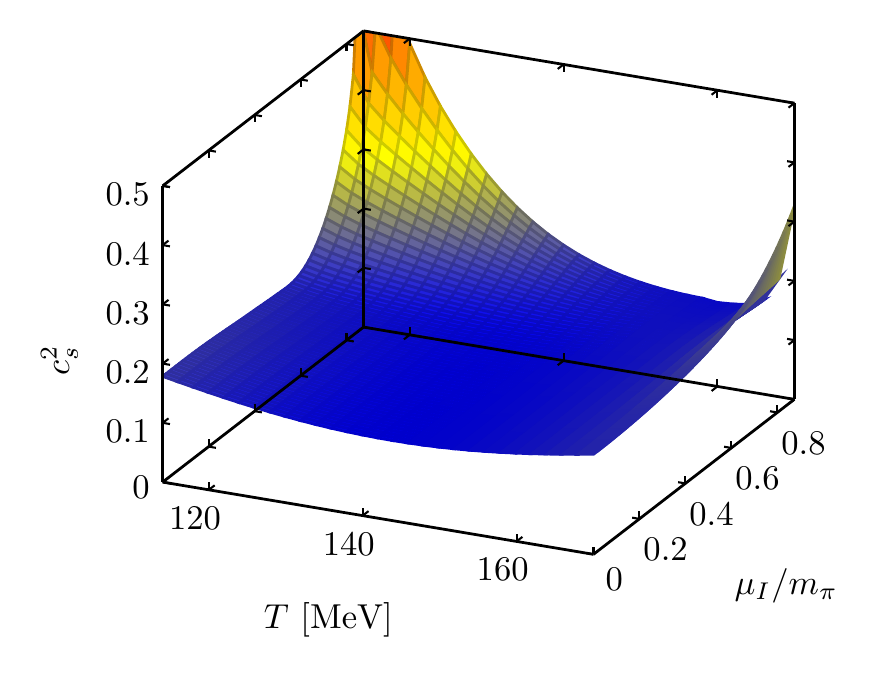} \\[-4.5mm]
 {\boldmath $N_t=10$}
 \end{minipage}
 \begin{minipage}{.48\textwidth}
 \centering
 \includegraphics[width=\textwidth]{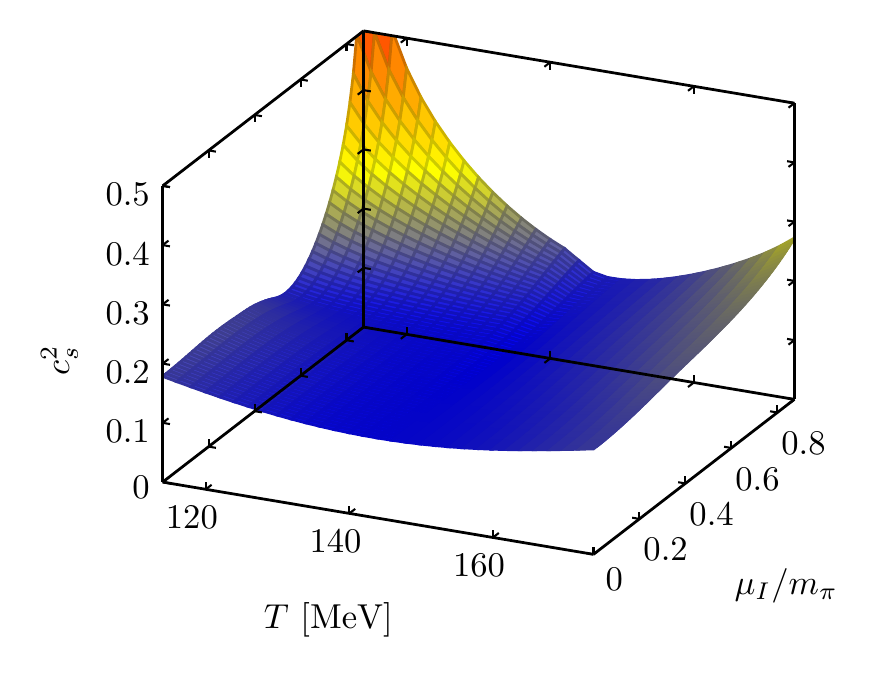} \\[-4.5mm]
 {\boldmath $N_t=8$} \\
 \includegraphics[width=\textwidth]{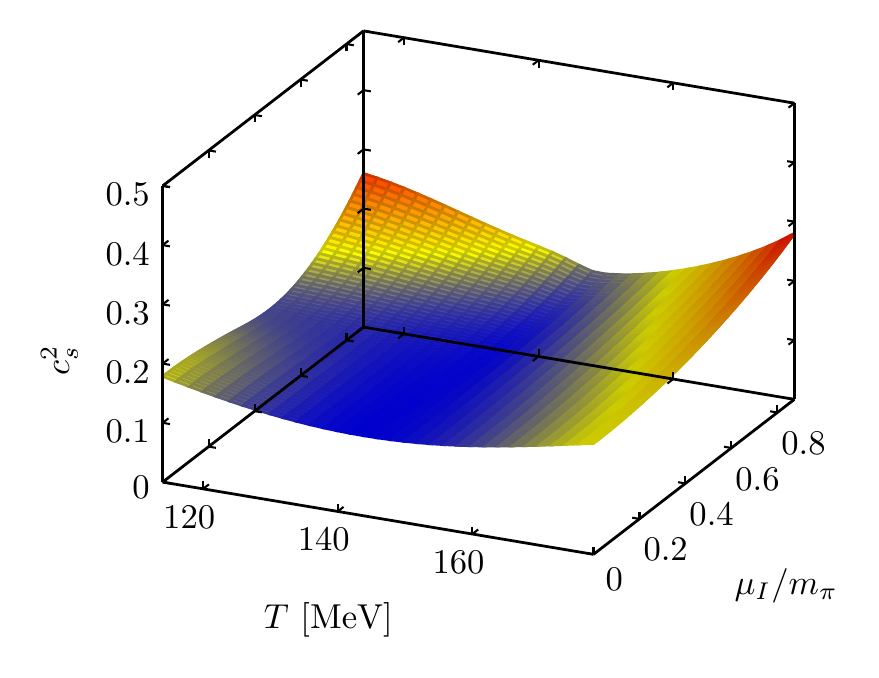} \\[-4.5mm]
 {\boldmath $N_t=12$}
 \end{minipage}
 \caption{
 Results for the squared speed of sound at $T=0$ (top left), with the same color coding as
 in Fig.~\ref{fig:eos-T0}, and at $T\neq0$ for $N_t=8$ (top right), 10 (bottom left)
 and 12 (bottom right). For the nonzero temperature panels, the uncertainties are again excluded for clearer plots.}
 \label{fig:cs}
\end{figure}

In our study we use the lattices of Ref.~\cite{Brandt:2017oyy}, to which we refer for further
details, which have already been used to map out the phase diagram up to $\mu_I/m_\pi\lesssim 0.9$.
In particular, we use the ensembles with $N_t=8,\,10$ and 12 and an aspect ratio $N_s/N_t\approx 3$.
From the isospin density we compute modifications of pressure and interaction measure and combine
this with the $\mu_I=0$ data obtained from the interpolation formula provided in
Ref.~\cite{Borsanyi:2013bia} with slightly modified coefficients, obtained from a reanalysis of
the data.\footnote{We thank K\'alm\'an Szab\'o for providing the coefficients and their correlations.}
The results for the EoS at $T\neq0$, in particular, the pressure (top left), the interaction measure
(top right), the energy density (bottom left) and the entropy density (bottom right), on the
$N_t=8$ ensemble are shown in Fig.~\ref{fig:eos-nt8}. Once more we can see the evidence for the
presence of the pion condensate in the EoS via the characteristic behavior of the interaction
measure at low temperatures. This behavior disappears for temperatures, where the
system remains outside the BEC phase even for large $\mu_I$. Similar results for the EoS
are also available for $N_t=10$ and 12, as shown in Ref.~\cite{Brandt:2021yhc}.

\section{The speed of sound}

From the above interpolations of the isospin density, one can also extract the isentropic
speed of sound $c_s$. which at non-zero isospin chemical potential is defined as
\begin{equation}
    c_s^2 = \left. \frac{\partial p}{\partial \epsilon} \right|_{\frac{s}{n_I}={\rm const}} = 
\frac{\partial_\xi\, p}{\partial_\xi\, \epsilon} \,.
\end{equation}
The latter defines the directional derivative in the direction of isentropes,
satisfying the condition
\begin{equation}
    \partial_\xi \left(\frac{s}{n_I}\right) = 0 \,.
\end{equation}
For $T=0$ $\delta_\xi=\delta_{\mu_I}$, while for $T\neq0$ the directional derivative mixes temperature- and chemical potential-derivatives. Given the
spline interpolations, all quantities can be computed analytically.

The results for $c_s^2$ at $T=0$ and $T\neq0$ are shown in Fig.~\ref{fig:cs}.
At $T=0$ (top left panel), the speed of sound increases strongly with $\mu_I$, crosses the
conformal bound~\cite{Cherman:2009tw} at $\mu_I/m_\pi\approx0.64$ and reaches a peak around
0.76 to 0.80. The value of $c_s^2$ at the peak
position is about 0.53 to 0.6. We note, that peak position and the height become
larger with decreasing lattice spacing, so that we expect these results to present lower
bounds to the continuum limit.
Our findings are in good agreement with recent results obtained in two-color
QCD~\cite{Iida:2022hyy,Itou:2022ebw}. For $T\neq0$, the remnant of this peak is
still visible for $N_t=8$ and
10, albeit shifted towards larger $\mu_I$ values. The tendency for $c_s^2$ to increase
is also still present for $N_t=12$, but we do not see the crossing of the
conformal bound up to $\mu_I/m_\pi=0.9$. This might be a sign that the peak at $T\neq0$ gets
shifted to even larger $\mu_I$ when approaching the continuum
limit. Note, however, that the peak appears at the outer sides of the spline
interpolations, where results are potentially more affected by systematic uncertainties.
For larger temperatures, $c_s$ increases and might even cross the conformal bound at
large $\mu_I$ values, but to see what is happening in this region further simulations
are necessary.

\section{Towards an extension to nonzero \texorpdfstring{\boldmath $\mu_L$}{muL} via Taylor expansion}

\begin{figure}[b]
 \vspace*{-2mm}
 \centering
 \includegraphics[width=.45\textwidth]{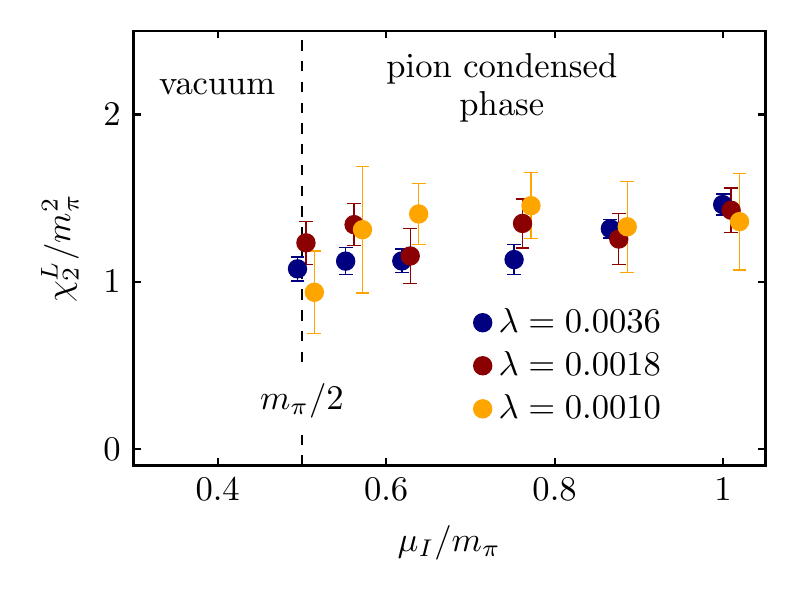}
 \caption{
 Results for the leading order Taylor coefficient $\chi^L_2$ in the BEC phase
 for different values of $\lambda$ (indicated in lattice units) obtained on our $T\approx0$ lattices at
 $a\approx0.15$~fm.
 }
 \label{fig:tcoeff-f1}
\end{figure}

We would now like to extend our determination of the EoS to non-zero but small light quark
chemical potentials $\mu_L$ from Eq.~(\ref{eq:chem-base}).
The extension can proceed via a Taylor expansion~\cite{Gottlieb:1988cq} of the pressure,
given by
\begin{equation}
    p(T,\mu_I,\mu_L)=\sum_{m=0}^\infty
   \frac{\chi^L_{m}(T,\mu_I)}{m!} \mu_L^m \qquad\text{with}\qquad
   \chi^L_{m}(T,\mu_I) = \left. \frac{\partial^{m}p(T,\mu_I,\mu_L)}
   {\partial \mu_L^m} \right|_{\mu_L=0} \,.
\end{equation}
The remaining task is to compute the Taylor coefficients for a given point in parameter
space $(T,\mu_I)$. Here we will focus on the leading order of the expansion, $m=2$.
The results obtained for different $\lambda$ of the Taylor coefficient $\chi^L_2$
on the $T=0$ lattices with $a\approx0.15$~fm are shown in Fig.~\ref{fig:tcoeff-f1}.
These results for the Taylor coefficients are unimproved with respect to the $\lambda$-extrapolations
and, at first glance, seem to facilitate a simple extrapolation to $\lambda=0$.

In Ref.~\cite{Brandt:2017oyy} we have detailed the improved $\lambda\to0$ extrapolation
procedure for condensates and densities. In the valence quark improvement we generically
split the operators associated with light quark fermionic observables by 
$\ev{O}=\ev{O-\delta^N_O} + \ev{\delta^N_O}$, with $\lim_{\lambda\to0} \ev{\delta^N_O} = 0$.
The improvement term $\delta^N_O$ is defined via an approximation of the trace appearing
in $O\sim\text{Tr}(\hat{O}\,M^{-1})$ in terms of singular values and the corresponding eigenstates
of the massive Dirac operator, $M^\dagger(\mu_I)M(\mu_I)\varphi_n=\xi_n^2 \varphi_n$, so that
\begin{equation}
 \label{eq:impro-term}
 \delta^N_O \sim \sum_{n=0}^{N-1} O_{nm} \Big(\frac{1}{\xi_n^2+\lambda^2} - \frac{1}{\xi_n^2}\Big) \,,
\end{equation}
where we have introduced the matrix elements of the operator $O$ with respect to the
eigenstates, $O_{nm}=\varphi^\dagger_n \hat{O} \varphi_m$.
Here this procedure has to be extended to susceptibilities. Those include traces with
two inverses of the Dirac operator, for which the contribution of low modes reads
\begin{equation}
 \ev{\text{Tr}\big(\hat{O}_1 M^{-1} \hat{O}_2 M^{-1} \big)} \approx \ev{\sum_{n,m=0}^{N-1} \frac{O_{1;nm}}{\xi_m^2+\lambda^2} \frac{O_{2;mn}}{\xi_n^2+\lambda^2}} \,,
\end{equation}
where the matrix elements are defined as above. Using this representation, we can
define the improvement term $\delta^N_{\chi}$ for the valence quark improvement of
a generic susceptibility $\chi$ analogous to Eq.~(\ref{eq:impro-term}) as
\begin{equation}
 \label{eq:lam-imp-susc}
 \delta^N_\chi \sim \sum_{n,m=0}^{N-1} O_{1;nm}O_{2;mn} \Big(\frac{1}{(\xi_n^2+\lambda^2)\,(\xi_m^2+\lambda^2)} - \frac{1}{\xi_n^2\,\xi_m^2} \Big) \,.
\end{equation}
In principle, one could truncate the two sums
at different numbers of singular values. This, however, is not beneficial numerically, since
the lowest singular value matrix elements are available immediately for all operators.
We note that the leading order reweighting of gauge configurations, as developed for condensates in Ref.~\cite{Brandt:2017oyy}, can be included in the same way.

\begin{figure}[t]
 \vspace*{-2mm}
 \centering
 \includegraphics[width=.45\textwidth]{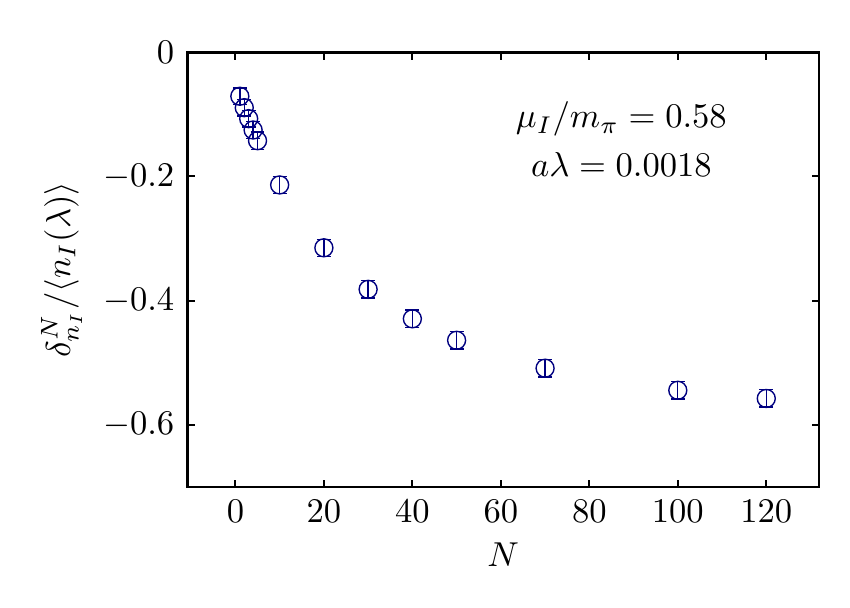}
 \includegraphics[width=.45\textwidth]{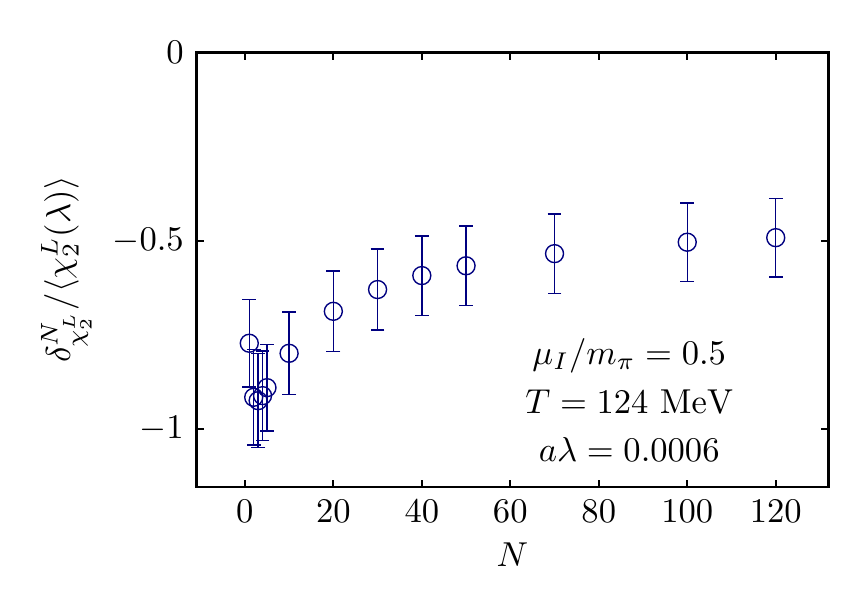}
 \caption{{\bf Left:} Results for the improvement term of the isospin density versus the number
 of singular values included in its computation on the $T=0$ lattice at
 $a\approx0.15$~fm with $\mu_I/m_\pi=0.58$.
 {\bf Right:} Results for the improvement term of the leading order Taylor coefficient
 $\chi^L_2$ versus the number  of singular values included in its computation
 on one of the $24^3\times6$ lattices with $T=124$ MeV.}
 \label{fig:tcoeff-f2}
\end{figure}

We show
the results for the improvement term $\delta^N_{n_I}$ for different values of $N$ in
the left panel of Fig.~\ref{fig:tcoeff-f2}. When we apply the same procedure for $\chi^L_2$,
however, the lowest singular values are found to dominate the improvement term more strongly, since they
appear with a higher power in the denominator of Eq.~(\ref{eq:lam-imp-susc}).
This has two effects: first, it leads to a larger correction term, which, however,
becomes visible in the unimproved susceptibilities only at smaller values of $\lambda$.
Hence the small dependence on $\lambda$ in Fig.~\ref{fig:tcoeff-f1}. Second, it increases
fluctuations in the correction terms and leads to larger uncertainties. This is visible
in the right panel of Fig.~\ref{fig:tcoeff-f2}, where we show the improvement term
$\delta^N_{\chi^L_2}$ versus $N$ for a nonzero temperature ensemble. We note, that the situation becomes worse at $T\approx 0$,
where the dominance of the lowest singular value is observed to be even stronger. This behavior
renders reliable $\lambda$-extrapolations of the Taylor
expansion coefficients more challenging and requires further optimization.

\section{Conclusions}

In this proceedings article we have presented the status of our study of the EoS of
isospin-asymmetric QCD at zero and non-zero temperatures. The EoS has been determined from a
model independent spline interpolation of the isospin density and at $T=0$ residual temperature
effects have been corrected using chiral perturbation theory~\cite{Son:2000xc} in the vicinity
of the transition to the BEC phase. In the BEC phase at small temperatures the interaction
measure shows a very distinctive feature: it initially increases until it reaches a maximum, followed by a reduction and eventually turns negative deep in the BEC phase.
Another, particularly interesting observable related to the EoS is the isentropic speed of sound.
At zero temperature, the speed of sound increases in the BEC phase, exceeds the conformal
bound~\cite{Cherman:2009tw} at $\mu_I/m_\pi\approx0.64$ and exhibits a peak around
$\mu_I/m_\pi\approx0.76$ to 0.80. The onset of this peak is still visible for small
non-zero temperatures on coarser lattices, but it becomes shifted towards larger $\mu_I$
values with increasing temperature and in the approach to the continuum limit. To our
knowledge this is the first time that a speed of sound larger than the conformal bound
has been observed in first principles QCD.
Finally, we discussed our preliminary results concerning the impact of small light quark baryon chemical
potentials $\mu_L$ (cf.\ Eq.~(\ref{eq:chem-base})) using Taylor expansion around the novel
expansion points at non-zero temperature and isospin chemical potentials. The determination of the corresponding susceptibilities in the zero pion source limit is observed to be more demanding than for simple quark bilinears like the pion condensate or the isospin density.
\\ \noindent

\noindent\textbf{Acknowledgements:}\\
This work has been supported by the Deutsche Forschungsgemeinschaft (DFG, German Research Foundation) via CRC TRR 211 – project number 315477589. The authors gratefully acknowledge the Gauss Centre for Supercomputing e.V. (\href{https://www.gauss-centre.eu}{\tt www.gauss-centre.eu}) for funding this project by providing computing time on the GCS Supercomputer SuperMUC-NG at Leibniz Supercomputing Centre (\href{https://www.lrz.de}{\tt www.lrz.de}).

\providecommand{\href}[2]{#2}\begingroup\raggedright\endgroup


\end{document}